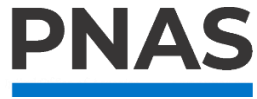

## Main Manuscript for

Specialized pro-resolving mediator Resolvin E1 corrects the altered cystic fibrosis nasal epithelium cilia beating dynamics.

Maëlle Briottet[1], Khadeeja Sy[1], Charlie London[1], Abdel Aissat[1], Mickael Shum[1,2], Virginie Escabasse[2], Bruno Louis[1] and Valérie Urbach[1]*

[1]INSERM U955- Institut Mondor de Recherche Biomédicale (IMRB), Créteil, France.

[2]Centre Hospitalier Intercommunal de Créteil (CHIC), France.

* Corresponding author: Valérie Urbach.

**Email:**  valerie.urbach@inserm.fr

**This PDF file includes:**

>Main Text
>Figures 1 to 6
>Tables 1 to 2

**Abstract**

In cystic fibrosis (CF), impaired mucociliary clearance leads to chronic infection and inflammation. However, cilia beating features in a CF altered environment, consisting of dehydrated airway surface liquid layer and abnormal mucus, has not been fully characterized. Furthermore, acute inflammation is normally followed by an active resolution phase requiring specialized pro-resolving lipid mediators (SPMs) and allowing return to homeostasis. However, altered SPMs biosynthesis have been reported in CF. Here, we explored cilia beating dynamics in CF airways primary cultures and its response to the SPMs, resolvin E1 (RvE1) and lipoxin B4 (LXB4). Human nasal epithelial cells (hNECs) from CF and non-CF donors were grown at air-liquid interface. The ciliary beat frequency, synchronization, orientation, and density were analyzed from high-speed video-microscopy using multiscale Differential Dynamic Microscopy algorithm and an in-house developed method. Mucins and ASL layer height were studied by RT-qPCR and confocal microscopy. Principal component analysis showed that CF and non-CF hNEC had distinct cilia beating phenotypes, which was mostly explained by differences in cilia beat organization rather than frequency. Exposure to RvE1 (10 nM) and to LXB4 (10 nM) restored a non-CF like cilia beating phenotype. Furthermore, RvE1 increased the airway surface liquid (ASL) layer height and reduced the mucin MUC5AC thickness. The calcium-activated chloride channel, TMEM16A was involved in RvE1 effect on cilia beating, hydration and mucus. Altogether, our results provide evidence for defective cilia beating in CF airway epithelium and a role of RvE1 and LXB4 to restore the main epithelial functions involved in the mucociliary clearance.

**Significance Statement**

Cilia beat dynamic controls the quality of the airway mucociliary clearance. In cystic fibrosis (CF), cilia beating features have not been fully characterized in a pathological environment including reduced airway surface liquid (ASL) layer and mucus abnormalities. Here, we developed a method using live cell microscopy on primary cultures of nasal epithelial cells from patients and considering the frequency, synchronisation, orientation, and quantity of cilia beating as well as the ASL height and secreted mucins. This analysis allowed to further describe a cilia dyskinesia in CF and provided evidence for the effect of the specialized pro-resolving lipid mediators, resolvin E1 in restoring cilia beating (figure 1). This multi-parametric method is a new tool to study motile cilia.

**Main Text**

**Introduction**

CF is a recessive monogenic disease caused by the mutation of the Cystic Fibrosis Transmembrane Conductance Regulator (CFTR) gene, encoding the CFTR chloride channel. More than 2000 mutations have been identified, but at least one copy of the F508del mutation, is expressed in 90% of the patients (1). Although, CF is a systemic disease, the airway disease is the main cause of morbidity and mortality. The loss of CFTR protein function leads to defective secretion of chloride and bicarbonate, and sodium hyperabsorption resulting in a dehydration of the airway surface liquid (ASL) layer that contributes to a viscous and sticky mucus and impaired mucociliary clearance. This altered ASL layer favors chronic infections and inflammation leading to progressive lung damage and respiratory failure (2). Inflammation response in CF, is described as sustained, exaggerated, and inefficient against infection (3–7).

Acute inflammation is normally followed by an active resolution phase to enable the return to tissue homeostasis. The resolution of inflammation requires the biosynthesis and functions of specialized pro-resolving mediators (SPMs) (8). SPMs such as lipoxins, resolvins, protectins and maresins exert numerous actions including the inhibition of the NF-kB pathway (10–12) and the inhibition of leucocytes migration (12). SPMs also stimulate the efferocytosis of apoptotic immune cells (13, 14). We and others have previously shown that lipoxin A4 (LXA4) levels were significantly reduced in the airway of patients with CF (15, 16). More recently, we demonstrated the contribution of CF nasal epithelial cells in the abnormal biosynthesis of LXA4 and several other SPMs (17). Therefore, lower levels of SPMs could contribute to the altered inflammatory response in the CF airway disease. SPMs treatment can improve various epithelial functions in models of CF airway disease (18). Previous works demonstrated that LXA4 and resolvin D1 (RvD1) were able to restore the ASL layer height in human bronchial epithelial cells (19-22), suggesting that SPMs could have an impact on the quality of mucociliary clearance. The roles of the two SPMs, resolvin E1 (RvE1, 5S,12R,18R-trihydroxy-6Z,8E,10E,14Z,16E-EPA) and lipoxin B4 (LXB4, 5S,14R,15S-trihydroxy-6E,8Z,10E,12E-eicosatetraenoic acid) have not been explored before on CF airway epithelial cell function (23,24).

The mucociliary clearance results from the transport of mucus out of the airways due to the cilia beating at the epithelial cell surface. However, the dynamics of cilia is not well established, especially in a pathological environment including the altered airway surface hydration and mucus content. Primary cultures of nasal epithelial cells at air-liquid interface (ALI) are useful models to mimic physiological environment, as we can obtain a fully differentiated pseudo-stratified epithelium with basal cells, goblet cells that produce mucus, and ciliated cells (17,25). In this model, CF primary cultures have shown a reduced and disrupted ASL layer height compared with primary cultures from non-CF donors (26). However, limited methods are available to assess cilia beating on such models. Most cilia beating analysis only measured cilia beating frequency (CBF). However, CBF alone cannot explain the capacity of cilia to propel the mucus layer. Other parameters such as amplitude, cilia beating pattern or coordination are measured in primary cilia dyskinesia studies. But these methods have mainly been used for imaging cilia beating from the side after removal of the cilia environment generated by epithelial cells (27-29). Cicuta's team recently developed an automated method, the multiscale Differential Dynamic Microscopy (multi-DDM) to obtain the CBF as well as the cilia synchronization of cilia beating in intact ALI cultures (30,31). This previous study demonstrated that the spatial and temporal coherence of cilia movement was altered in CF samples. Another study developed on ALI cultures characterized a third parameter of cilia movement influenced by mucus environment: the cilia beating orientation (32).

In our work, we aimed to characterize the dynamics of cilia beating on a model mimicking the pathophysiological context of CF airway disease and by considering as many parameters as possible, including the cilia beat frequency (CBF), synchronization, orientation, density as well as the temperature which is a well-known modulator of cilia beating and mucociliary clearance *in vitro* and *in vivo* (29,33). In order to identify the level of dependency of each parameter and guide their interpretation, we performed a principal component analysis (PCA) which allowed to further characterize a differential cilia beating phenotype in nasal epithelial primary cultures derived from CF patients compared to control subjects (figure 1). Moreover, we investigated the impact of RvE1 and LXB4 on cilia beating in CF and the cellular mechanisms involved.

**Results**

**Characterization of cilia beat phenotype of airway epithelial primary cultures using a multiparametric analysis**. We performed a PCA for five quantitative parameters: the density of cilia in movement (movement), the cilia beat frequency (CBF), the synchronization determined by the surface size ( $\lambda^2$ ) below which cilia beat in a homogeneous way, the orientation (the dispersion of cilia beat orientation over the video) and the room temperature during acquisitions (figure 2A). The correlation circles showed that principal components (PC) 1 to 4 explained 33.1 %, 26.21%, 18.80% and 14.98% of the total variance, respectively. Therefore, the first 2 PC could explain 59.31% of variance of the data. The CBF and temperature were the two main contributors of variance for the first PC. Percentage of movement, synchronization ($\lambda^2$) and orientation of cilia beating appeared rather independent of the CBF and mainly contributes to the second PC (figure 2A, left). The density of movement negatively correlated with the orientation and synchronization

of the cilia beating. These last two parameters positively correlate over the 2nd PC. In contrast, cilia beat orientation and synchronization appeared negatively correlated over the 3rd PC, suggesting that they contributed differently to PC2 and PC3 (figure 2A). The synchronization ($\lambda^2$), participated in 35.45% of the variance explained by PC2 and 32.86% for PC3, whereas orientation participated in 19% for PC2 and 64.3% in PC3. This suggested that synchronization and orientation of the beating could bring complementary information. Taken together, the PCA has shown that the percentage of movement, the cilia beat frequency, synchronization and orientation gave distinct information that could enable a fine characterization of cilia beating phenotype.

**Distinct cilia beating phenotypes between non-CF and CF samples.** We compared the cilia beating phenotypes between CF and non-CF untreated samples and observed different profiles which are visualized by the ellipses for the two first components (figure 2B and supplemental figure 1 for PC3 and PC4). At 24h, 48h and 72h after washing the mucus, the ellipses from non-CF and CF datasets were not stackable and were oriented in opposite directions (figure 2B). 24h after removing the mucus, the centroids of cilia beating data from non-CF and CF groups were significantly different (p=0.001) while the dispersion of the was similar (p= 0.531). The differences between the CF and non-CF groups were still observed after 48h and 72h. The cilia beating phenotype for the F508del homozygous samples only was not significantly different from the CF group (4 F508del homozygous and 2 heterozygous) (figure 2B, dashed line). The temperature gradient over the scatter plot illustrates how temperature changed along the first PC and that the non-CF and CF experimental sets differentiated more at higher temperature (red) than at lower temperature (yellow) and mostly along the PC2 axis (figure 2B). Since cilia beat density, synchronization ($\lambda^2$) and orientation were the major contributors to the PC2 axis, these parameters were mainly responsible for the differences between CF and non-CF samples rather than the CBF. This analysis was confirmed when the four parameters were analyzed separately. The CBF values did not show differences between CF and non-CF samples 48h and 72h after mucus removal (figure 2C). The CBF slightly increased in CF hNEC 24h after mucus removal which is consistent with the slight but significant increase in the temperature at 24h (WT: 23.19°C; CF: 23.94°C; p<0.01) since a high correlation between CBF and temperature was observed. In contrast, the percentage of movement, the synchronization ($\lambda^2$) and orientation of cilia beating were significantly different between CF and non-CF samples. CF samples showed a lower degree of cilia movement for the three time points after removal of mucus (figure 2D). In CF hNEC, the cilia beating was coordinated over a larger area than in non-CF samples (higher $\lambda^2$) (figure 2E). Finally, the cilia beat orientation showed a higher dispersion for CF samples compared with non-CF samples (Figure 2F).

To consider the impact of the temperature revealed by the PCA, we further compared the beating parameters within two subgroups of lower ([19°C-24°C]) and higher (]24°C-28°C]) room temperature, 24h after mucus removal (supplemental figure 2). This confirmed that at the lower temperatures, the beating parameter were not significantly different between CF and non-CF except for the orientation (higher dispersion in CF). At higher temperatures, all the parameters showed significant differences between CF and non-CF samples with decreased CBF and density of movement and increased synchronization and dispersion of cilia in CF. The PCA analysis also showed that the density was not much correlated to the CBF but appeared inversely correlated to synchronization and cilia orientation. Therefore, to minimize the influence of the quantity of cilia in movement to the other parameters, we tested the restriction of the analysis to the videos containing over 50% of movement in both groups (supplemental figure 3). This resulted in reduced differences of movement between CF and non-CF samples while keeping enough videos. However, despite the restriction, significant differences in cilia synchronization and orientation between CF and non-CF samples were kept (supplemental figure 3). This suggested that the differences in cilia synchronization and orientation between CF and non-CF did not only result from the reduced number of cilia in movement in CF culture.

After mucus removal, cilia beat synchronization and orientation did not evolve much over time within the same sample's group (CF or non-CF) (figure 2 and table 2). In contrast, the CBF decreased after mucus removal, especially for the acquisition performed at the higher temperature (]24°C-28°C]) the CBF dropped by 1.7Hz (p<0.0001) and by 1.2Hz (p<0.0001) between 24 and 72h in non-CF and CF samples respectively. For lower acquisition temperatures the CBF for non-CF samples did not change and dropped by 1Hz (p<0.01) in CF samples.

Taken together, the cilia beating characteristics of CF and non-CF hNEC were significantly different at any time after mucus removal which is consistent with an impaired cilia beating in hNEC from CF samples. More specifically, while we did not detect significant CBF differences, we showed a reduced cilia density, a higher synchronization and dispersion of cilia orientation in CF hNEC. These differences could at least partially result from the changes in cilia environment between CF and non-CF samples, CF environment restricting the beating orientation and synchronization.

**SPMs regulate cilia beating of CF airway epithelium.** We then explored whether SPMs could regulate cilia beating parameters. The CF hNEC, were treated 24h after mucus removal with SPMs (10nM) for 1h, 24h or 48h. Therefore, the samples exposed to SPMs for 24h, accumulated mucus for 48h, and similarly the ones exposed to SPMs for 48h which accumulated mucus for 72h. Figure 3 displays the effects of LXB4 and RvE1 tested on CF hNEC. For each condition we compared CF samples treated with SPMs (in green) to CF control samples treated with the vehicle (in blue), and to non-CF control samples (treated with the vehicle, in grey). We found that LXB4 (10nM) treatment significantly moved the ellipse from the CF beating phenotype toward a non-CF phenotype after 1h and 48h treatments (25h and 72h after mucus removal). RvE1 treatment of hNEC cultures from 5 different patients also tended to modify the cilia beating of CF samples towards a non-CF profile, which was significant at 24h treatment.

A further analysis of the RvE1 effects on hNEC from each CF donor separately revealed that responses to RvE1 were donor specific even between donors carrying the same CFTR mutation. The impact of RvE1 on hNEC cultures from two different patients, both homozygous for the F508del mutation CF1 (n=3) and CF2 (n=2) are presented in the figure 4. For the hNEC derived from the first CF patient (CF1), the PCA analysis showed that the RvE1 treatment significantly restored the cilia beating phenotype towards a non-CF phenotype starting from 1h and increasing to 48h post treatment. After 24h and 48h, the beating pattern of the treated samples overlays the non-CF cilia beating pattern (figure 4A). Figure 4B, showed that RvE1 restored cilia beating frequency (increased), and dispersion of cilia orientation (decrease) and tended to restore synchronization (decreased) in hNEC-CF1. LXB4 treatment on hNEC from this patient (CF1) also resulted in improving the cilia beating (Supplemental figure 4). In contrast, for hNEC cultures derived from the other patient CF2, the PCA analysis as well as the separate analysis of cilia beating parameters showed that RvE1 failed to modify any cilia beating parameters (Figure 4C and 4D). The comparison of the initial beating profiles of the hNEC from these 2 different patients (CF1 and CF2) showed that they were not similar. Indeed, before treatment, the hNEC that responded to RvE1 (hNEC-CF1), were synchronized over a higher surface of 368µm² and 291µm², with a dispersion of orientation of 54 and 51 degrees, after 48h and 72h of mucus accumulation, respectively. In contrast, the hNEC that did not respond to RvE1 (hNEC-CF2) were synchronized over a smaller surface of 150µm² and 268µm², with a dispersion of orientation of 45.4 and 48 degrees. These last values were close to the values obtained in non-CF samples with a cilia synchronization of 170µm² and 200µm², and a dispersion of orientation of 46 and 47.5 degrees.

**The role of TMEM16A in the effect of RvE1 on airway surface hydration and cilia beating.**
We studied whether the effect of RvE1 on cilia beating profile could be related to modifications of the cilia environment. Since the CF airway surface dehydration results in a more viscous mucus, we investigated the effect of RvE1 on the hydration of the surface of hNEC. The ASL layer height was visualized using live cell confocal microscopy upon treatment with RvE1 (10nM) or vehicle control. Figure 5A and 5B, showed that in control conditions, the ASL was thin (4.4 ± 0.09 µm) and disrupted. The treatment with RvE1 for 1h significantly enhanced an ASL layer height increase to 8.2 ± 0.35 µm. The role of TMEM16A, a calcium-dependent chloride channel, was investigated by using its specific inhibitor Ani9. Exposure to Ani9 (10µM) completely inhibited the RvE1 effect in increasing the ASL layer height, suggesting that RvE1 restored the ASL layer by stimulating the TMEM16A chloride channel. We then investigated the effect of inhibiting TMEM16A with Ani9 on the cilia beating. The effect of Ani9 was illustrated by the changes in ellipses orientation of the PCA analysis towards a CF phenotype (without SPM) rather than the non-CF phenotype induced by RvE1 (Figure 5C). Ani9 also inhibited the effect of RvE1 on CBF and on cilia orientation especially at 48h (Figure 5D). These results suggest that the impact of RvE1 on cilia beating parameters involves at least partially the effect of this SPM in restoring airway epithelium surface hydration through the stimulation of calcium-dependent chloride secretion by the TMEM16A channel.

**RvE1 modifies secreted mucins organization.** To evaluate the possible role of mucins in the differences of cilia beat profile between non-CF and CF hNEC and upon RvE1 treatment, we explored the secretion and transcription of the two secreted mucins MUC5B and MUC5AC. Immunofluorescence labelling of the secreted mucins by hNEC from the CF1 patient and by non-CF samples 72h after mucus removal are displayed in figure 6A. MUC5B (in red) and MUC5AC (in green) partially colocalized and constructed a mesh and different structures could be observed. In non-CF primary cultures, the mucins adopted a structure of linear strands or thin packed bundles. In contrast, in hNEC CF1 primary cultures, the mucins stained thicker than in non-CF cultures. They were more structured as sheets than linear strands (figure 6A). Furthermore, the thickness measurement of the MUC5AC staining showed a thicker mucus layer in CF1 hNEC sample (4.1 ± 0.11 µm) than in the non-CF samples (3.1 ± 0.05µm) ($p<0.0001$) (figure 6B).

The RvE1 exposure of CF1 hNEC for 48h induced a mucins' structure change with more linear strands than in control (vehicle). In addition, this treatment decreased MUC5AC thickness of hNEC toward non-CF values (3.5 ± 0.08 µm) (figure 6A and 6B). Finally, exposure to Ani9 prevented the effect of RvE1 in modifying MUC5AC structure and the mucin thickness was significantly increased (4.8 ± 0.11µm) compared with the condition where the cultures were only exposed to RvE1 (figure 6A and 6B).

At the transcriptional level, MUC5AC and MUC5B levels were higher in CF samples, although the differences only reached significance for MUC5AC. RvE1 (10nM) treatment during 24h did not significantly affect either MUC5B or MUC5AC mean expression level in CF samples (figure 6C). Thus, this further analysis was consistent with the assumption that CF airway epithelia were covered with a thicker mucins layer than non-CF cells and that RvE1 treatment regulated it.

**Discussion**

This biophysical study provided a global picture of multiple parameters of cilia beating that are altered in CF nasal epithelial primary culture. Moreover, this is the first report showing that cilia beating orientation was more dispersed in CF than in non-CF nasal epithelial cells primary cultures. Our data also demonstrated the role of the SPMs, RvE1 and LXB4 in restoring at least partially, cilia beating in CF nasal epithelial cells and that RvE1 effect involved the calcium-activated chloride channel, TMEM16A.

**Decrease of cilia beating quantity in CF airway epithelial cells.** Several papers reported a difference in the number of ciliated cells between CF and non-CF airway epithelial cells. Cell transcriptomic of proximal airway of CF donors at end-stage lung disease displayed a higher proportion of ciliated cells. However, when the samples were cultured at ALI, a reduced quantity of ciliated cells were observed in the CF samples (34). Another study reported a decrease in ciliated cells staining in CF lung tissue and primary cultures (35). A more recent analysis showed that 6 weeks of hNEC cultures at ALI, resulted in differences in ciliated cells differentiation between CF and non-CF donors (36). In our study, we did not strictly quantify the number of ciliated cells, we used instead an automated method to quantify the percentage of movement on each video recorded and we showed that this parameter significantly decreased by >20% in CF samples compared with non-CF ones. The PCA analysis also revealed that the percentage of movement was one of the parameters that discriminated most CF and non-CF samples. For each culture insert, the acquisitions were performed at 6 to 8 different locations, each time trying to maximize the quantity of cilia moving in the region of interest. Therefore, although it cannot be excluded that the cilia movement might have been more difficult to detect in CF samples, our results are consistent with previous reports showing that CF cultures were less densely ciliated.

**No change in cilia beat frequency in CF airway epithelial cells.** Previous studies performed on CBF in CF airway epithelial cells gave contradictory results. A normal CBF of CF nasal brushing *in vitro* was initially reported (37). In another study, the CBF in CF nasal biopsies and airway epithelial primary culture at ALI was found higher than healthy controls (38). In contrast, others reported that CF nasal brushings or ALI airway epithelial primary cultures showed significantly lower CBF and dyskinesia compared with that of healthy subjects (39, 40,41). Our data are consistent with a more recent study performed with similar tools as the ones we used (ALI primary cultures and multi-DDM method), the CBF ranged from 2.4 to 5.1 Hz with no perceptible differences between CF and non-CF samples (30). Furthermore, our PCA also suggested that CBF was not the parameter mainly responsible for the differences between CF and non-CF cilia beat profiles, highlighting the interest of exploring other parameters.

**Spatial homogeneity of cilia beat frequency increased in CF airway epithelial cells.** We explored the homogeneity of cilia beat by calculating the surface size in which cilia beat at similar frequencies and determine the level of cilia coordination by the $\lambda^2$ value (in $\mu m^2$) which is accountable for the spatial and temporal organization of cilia beating. Our data obtained on nasal epithelial cells are consistent with previous measurements of this parameter in CF and non-CF bronchial epithelial cell primary cultures showing that, 24h after mucus removal, the $\lambda^2$ in CF samples was higher than in non-CF samples (30). Since epithelial cells apical surface are between 10 to 100µm2, these results indicated that in CF, cilia beating was homogeneous over 3 to 30 cells and in non-CF cells, cilia beating was homogeneous only over 2 to 10 cells (depending on cell size). Consistent with this individual analysis, the PCA revealed that the synchronization level was also one of the main parameters that discriminated CF and non-CF cilia beat patterns. We also showed that the percentage of movement appeared inversely correlated to synchronization, while it was not much correlated to the CBF. When we are restricting the analysis to videos containing over 50% of movement, the differences of synchronization between CF and non-CF samples were then less significant. This result was consistent with previous studies showing that cilia density influences their organization (32).

**Dispersion of cilia beat orientation increased in CF airway epithelial cells.** We introduced a fourth parameter to consider the diversity of cilia beating orientation which has not been previously studied in CF. Our method permitted the calculation of the main cilia direction over square boxes of 16.7µm. Therefore, we looked at cilia main beating orientation in area ~280µm². We thus studied the dispersion of cilia orientation between areas were CF and non-CF beating were no more correlated ($\lambda^2$ = 280µm2). This suggests that the orientation parameters gave other information than $\lambda^2$; these two parameters being related to 2 different spatial scales of analyses. Furthermore, the PCA indicated that $\lambda^2$ and orientation contributed differently to PC2 and PC3 which was also consistent with our hypothesis that cilia beating orientation is a complementary characteristic of cilia dynamics. Our data provided the first evidence that orientation of cilia beating was more dispersed in CF than in non-CF nasal epithelial cells primary cultures while cilia beating (mainly CBF) was more homogenous over bigger areas for CF samples. In addition, when restricting the analysis to acquisitions where >50% of movement was detected, the dispersion in cilia orientation in CF samples detached further from non-CF samples suggesting that orientation was more independent of the percentage of movement than the $\lambda^2$ (the area within cilia beat at the same frequency). The higher dispersion of cilia orientation in CF cells compared with non-CF could be consistent with a recent study showing that sheets of mucus over 200µm² covering the cell surfaces and compressing cilia in CF cells resulted in disorganized and flattened cilia (42).

**Temperature's increase stimulated cilia beat in CF airway epithelial cells.** Several studies have shown that increasing temperature had a positive effect on mucociliary clearance. The mucociliary clearance rates of nasal fluid measured *in vivo* was higher at 37°C than 20°C (33). The mucociliary clearance measured *in vitro* on tracheal ovine model was also reduced at lower temperature (43). In primary cilia dyskinesia, cilia beat pattern is more easily observed as dyskinetic at 22-24°C than at 4°C when measured on brushed cells (29). In these studies, whether the effect of temperature on mucociliary clearance was only due to improvement of CBF or also due to modifications of cilia coordination was not clear. However, a linear positive dependence of CBF with temperature has been reported, notably between 19°C and 32°C, when measured on nasal epithelial brushings (44). Consistent with this previous data, our study showed a strong positive correlation between temperature and CBF and that increasing temperature (from 20°C to 28°C) stimulated CBF significantly ($p<0.0001$) in both CF and non-CF hNEC. Temperature also had an impact on the other parameters of cilia beating. Higher temperatures increased the density of movement of non-CF hNEC but decreased it in CF hNEC. Increasing temperatures significantly decreased synchronization and orientation in non-CF hNEC but had less effect on CF cells.

Our study also revealed that at the higher working temperature (24°C-28°C), CF samples showed a lower CBF and density of movement and a higher synchronization and orientation than non-CF ones. In contrast, at the lower temperatures (19°C-24°C) cilia beating phenotype for non-CF and CF cultures were more alike. It has been previously reported that decreasing temperature from 37°C to 29°C for a couple of hours, stimulated F508del CFTR trafficking to the plasma membrane resulting in the rescue of chloride transport (45,46). Although in our experiments the epithelial cells moved from 37°C to room temperature for 10-15 min for the acquisitions, it cannot be excluded that a CFTR rescuing process at lower temperature could explain the lack of cilia beating differences between CF and non-CF samples. Taken together, our data suggested that acquisitions performed at body temperature could reveal larger cilia beating differences between CF and non-CF epithelial cells.

**Mucus accumulation altered the cilia beat in CF epithelial cells.** Early studies reported a normal CBF of cilia from CF nasal brushing *in vitro* together with an abnormal nasal mucociliary clearance measured *in vivo* in the same patients (37). This suggested the role of mucus abnormalities in the altered mucociliary clearance. Indeed, in this previous study, CBF was analyzed on cells resuspended in an aqueous solution that diluted the mucus and facilitated cilia beating (37). Organization of cilia beating over long distances has been explained by hydrodynamic coupling (32, 47-50). In numeric modelling viscosity of cilia environment has been found to modulate their organization (32). In another model, it was shown that cilia are only synchronized locally and that the size of local synchronization domains increases with the viscosity of the surrounding medium (51). Therefore, the differences of CBF between CF and non-CF measured in our model, where mucus and ASL layer hydration spontaneously evolved are consistent with the role of mucus concentration on CBF, as previously reported. In addition, we observed that CBF values decreased overtime with mucus accumulation and that the CBF differences between CF and non-CF samples were no longer detectable at later stages, suggesting that mucus accumulation was a strong limiting factor of the CBF. In contrast, the size of the area in which CBF appeared homogenous ($\lambda^2$) did not evolve much over time during mucus accumulation. This suggested that 24h was enough to secrete the amount of mucus necessary to constrain the spatial organization of cilia beating. Further mucus accumulation did not modify the spatial organization of cilia movements while this accumulation contributed to slow down the beating (decreased CBF). Cilia orientation did not evolve much neither during mucus accumulation. Recent reports indicated that orientation of cilia beat is not predefined and that that cilia can reorient from 35° to 40° in 24h to align to the mucus flow while when mucus is removed, their direction is more dispersed (52, 32). In addition, a few studies reported that in healthy conditions cilia are not perfectly aligned and that small heterogeneity in cilia orientation optimized mucus transport (52, 53).

In our study, we observed a thicker mucus layer above CF cells and increased MUC5AC transcription level which were consistent with previous reports showing a hyper-concentration of mucins, especially MUC5B and MUC5AC in CF airways resulting in a compressed periciliary layer (35, 54–56, 59). Moreover, in CF tracheobronchial epithelial cells, mucus structure is also altered, and defective maturation of mucins after their secretion leads to a compact structure instead of a linear one (57). In BALs from young CF patients, bigger and less soluble flakes, made of densely packed mucins MUC5B and MUC5AC were described (58). These changes in mucus concentration and structure have been mainly related to the altered airway hydration because of ion transport defect in CF.

**SPMs restored airway surface hydration, mucus organization and cilia beat.** Before our study, the effect of SPMs on cilia beating had not been explored. We found that RvE1 and LXB4 were able to restore CF cilia beating parameters, in CF samples where cilia beating was impaired. Furthermore, the effect of RvE1 in enhancing the ASL height and decreasing the thickness of mucus was consistent with an improvement of cilia beating. Indeed, proper hydration determines the quality of the mucociliary clearance. The hydration maintains a distinct mucus and periciliary layer which are necessary for efficient mucociliary clearance (60, 61). The inhibitory effect of Ani9 on the RvE1 effect on hydration, on mucus thickness and on cilia beating, suggested that these effects were due to the stimulation of the TMEM16A channel. TMEM16A is a calcium-dependent chloride channel which has been considered as a therapeutic target to bypass the loss of chloride secretion by CFTR in CF. It was reported that TMEM16A expression is diminished in CF and that increase of TMEM16A expression, and stimulation improved chloride secretion and mucus transport rate (62,63). However, whether TMEM16A should be activated has been controverted. Indeed, the TMEM16A contribution to chloride secretion was found to be marginal and an increase in TMEM16A in goblet cells and mucus production have been observed during inflammation responses (64). On another hand, TMEM16A expression in goblet cells could also be beneficial to hydrate newly secreted mucins (65). The simultaneous effect of RvE1 in hydration, in decreasing mucus thickness and restoring cilia beat is consistent with a beneficial role of TMEM16A. Changes in mucins structure in CF have been linked principally to airway hydration and to a lesser extent extracellular pH and bicarbonate concentration (57,67). Bicarbonate has been shown to play a role in mucin expansion (67-70). TMEM16A is permeable to bicarbonate (71, 72) and was found to regulate goblets cells exocytosis (73, 74).
This novel effect of RvE1 at nanomolar concentration in restoring cilia beating parameters by increasing ASL height and reducing the thickness of the mucus layer is consistent with its role in airway innate immunity reported in other models (75–77). Indeed, RvE1 improves lung inflammation, airway mucus and airway hyper-reactivity in a murine model of asthma (76). RvE1 promotes oral mucosal surface clearance of neutrophils via its epithelial receptor ChemR23 (78). In our study, we did not explore the role of ChemR23 in the effect of RvE1 on cilia beating phenotype. However, ChemR23 was found highly expressed in the lung and its role in damping inflammation has been reported in this tissue (79,80). At higher concentrations, RvE1 effects can be mediated by BLT1, which is also expressed in human airway epithelial cells (81,82). Taken together our finding could open novel therapeutic strategies for CF patients that are not eligible or do not respond to CFTR modulators.

**Benefit of the method of multiparametric analysis developed.**
Our method developed to observe cilia beating in their pathophysiological environment, including periciliary layer and mucus permitted to study five parameters related to cilia beating using an automatic analysis of video microscopy records. The PCA allowed to determine the dependency level of each parameter. The PCA and the separate analysis for each parameter were consistent with each other. However, the PCA analysis appeared to be more sensitive and allowed to reveal significant differences of the cilia beat profiles between non-CF and CF samples as well as the impact of SPMs. Such analyses might be useful to study other diseases of the airway track or related to other organs with motile cilia and to screen the patient dependent responses to drugs.

**Conclusion**
This multi-parametric study further described the alteration of cilia beating in CF. Furthermore, we provided evidence for the role of RvE1 and LXB4 in restoring cilia beating, with RvE1 changing mucins spatial organization and enhancing ASL layer hydration through TMEM16A channel stimulation. The method also allowed to identify responsive and non-responsive donors' nasal epithelial cultures despite their same CFTR mutation.

## Materials and Methods

### Study population
Nasal polyp or mucosa samples from nine non-CF and six CF donors of the *Centre Hospitalier Intercommunal de Créteil* (CHIC, France) were excised after informed consent (Table 1). All experiments were performed in accordance with the Declaration of Helsinki and the Huriet-Serusclat and Jardet law on human research ethics. The study of these human samples was approved by the *Comité de Protection des Personnes Ile de France* (ID-RCB: 2011A00118-33).

### Primary culture
The human nasal epithelial cells (hNEC) from patients were cultured under ALI in Pneumacult medium (StemCell) as previously described (17). The apical mucus secretions were washed twice a week by gently adding 200µL of warm medium from the basolateral compartment for one minute. Furthermore, to study cilia beating over mucus accumulation, the mucus was washed rigorously 24h before the first video acquisitions (83) (Supplemental for details).

### High speed video microscopy acquisitions
Videos of cilia were acquired from above using an inverted Axiovert 200M microscope (Carl Zeiss) at magnification 40x (NEOFLUAR 40x/0.75) and a high-speed video camera PL-A741 (PixeLink). The videos of 400 x 400 pixels were recorded at 158fps, for 10s (28). The cilia beating was acquired in 6 to 8 different locations of the insert. The inserts were kept at 37°C between acquisitions. The SPM (10nM) or vehicle solution were applied on the basolateral side. For every culture insert, videos were taken before SPM's treatment and after 1, 4, 24 and 48 hours. Ani9 (10µM) was applied 30min before SPM's treatment (Supplemental for details).

### Calculation of cilia beating parameters
The density of cilia beating was evaluated using the standard deviation projection of the video and considering all pixels above a threshold of 30, for 8-bit encoded images. The frequency and the coordination were calculated using multiscale Differential Dynamic Microscopy (multi-DDM) (30). The coordination described as the surface ($\lambda^2$ in µm²) below which cilia beat in a homogeneous way was calculated by dividing the video in square boxes ranging from 2.7 µm to 133.6µm. The orientation was calculated as the standard deviation of principal beating direction over square boxes of 50 pixels (17 µm) (Supplemental for details).

### Multiparametric analysis using Principal Component Analysis
The PCA was calculated over 1659 videos recorded for untreated non-CF samples to be compared with untreated or SPMs treated CF samples, using the Stats package in R. Videos were taken at different time points up to 72h after thorough mucus washing. We analyzed five quantitative parameters: the density of movement (Movement), the temperature of acquisition (Temperature), the cilia beating frequency (CBF), the synchronization of cilia beating ($\lambda^2$), the cilia orientation (Orientation).

### ASL layer height analysis
The ASL layer height was measured as previously described using Texas red®-dextran (10,000MW, Invitrogen, Auckland, NZ) (19). Epithelial cells were stained using Calcein-AM (Invitrogen, Auckland, New Zealand). Epithelia were treated in the basolateral compartment with RvE1 (10nM) for 1 hour. Z-scans of the epithelia were acquired with a Zeiss LSM 510 Meta (40X). For each culture insert, the ASL height was measured at 27 different locations.

### Secreted mucins visualization
Culture inserts used for video-microscopy were fixed with Carnoy solution and permeabilized. Mouse monoclonal anti-MUC5AC (Abcam ab212636) 1:1000, rabbit polyclonal anti-MUC5B (Thermofisher PA5-82342) 1: 500, rat monoclonal anti-α-tubulin (YL1/2) (Thermofisher MA1-80017) 1:500 were used as primary antibodies. Images were acquired using an inverted confocal microscope (Zeiss Axio Observer 7 with LSM 900 Airyscan 2, 20x objective) and analyzed using FIJI. For each acquisition, 50 measures of the MUC5AC layer thickness were taken.

**Mucin transcripts quantification**

Mucin transcripts were evaluated with RT-qPCR on CF and non-CF samples. The cDNA sequence amplification was achieved using Taqman Master Mix (Thermofisher 4444557) and the primers for mucins MUC5B (Hs00861595_m1), MUC5AC (Hs01365616_m1) and the housekeeping gene beta-2-microglobulin (Hs00984230_m1).

**Reagents**

LXB4 was purchased from Cayman (90420) and RvE1 from VINRESOL Co (Budapest, Hungary, 20010207), kept at -80°C and used at 10nM. Ani9 (Sigma, SML1813) was used at 10µM.

**Statistical analysis**

Number of samples used are indicated in Table 1. The centroids of the PCA "clusters" and data dispersion were compared using the PERMANOVA test (84) and PERMDISP test (85) respectively from the vegan package on R. On bar graphs, data are represented as mean values ± SEM. Statistical significances were tested using an unpaired t-test or an equivalent non-parametric test, Mann-Whitney, using GraphPad InStat software. Statistical significance was defined as $p < 0.05$.

**Acknowledgments**

The authors thank the imaging platform of the *Institut Mondor de Recherche Médicale* (Créteil, France) and the support of X. Decrouy. This study was funded by the French National Institute of Health (INSERM), the CF patient association *Vaincre la Mucoviscidose* and the doctoral school of Life Science and Health in Créteil (France).

**Legends**

**Figure 1. Graphical abstract of the study.** We developed a new method of multiparametric signal analysis of cilia beating on air-liquid interface primary cultures of nasal epithelial cells from patients with cystic fibrosis (CF). The combined study of the frequency, synchronisation, orientation and quantity of cilia beating as well as of the ASL height and secreted mucins allowed to describe a cilia dyskinesia in CF that can be corrected by the specialized pro-resolving lipid mediators, resolvin E1.

**Figure 2. Distinct cilia beating phenotypes of CF and non-CF hNEC.** The principal component analysis (PCA) was performed over 1659 videos with 5 quantitative parameters: the percentage of movement (Movement), the temperature of acquisition (T°C), the cilia beat frequency (CBF), synchronization ($\lambda^2$) and orientation (Orientation). **A.** The correlation circles are displayed for the 4 first PC. **B.** Cilia beating from CF and non-CF cultures (treated with vehicle), were recorded at different time after washing out the mucus: at 24h (116 video of non-CF and 272 of CF hNEC), at 48h (120 non-CF and 75 CF) and 72h (91 non-CF and 42 CF). Scatter plot on the first two PC is displayed. Ellipses were drawn using stat_ellipse function (package ggplot2) and calculated using multivariate Student distribution and 95% confidence interval. PERMANOVA statistical tests were performed to compare cilia beat phenotype (centroid) recorded on CF (blue) and non-CF hNEC (grey). Ellipses for the F508del homozygous samples only are also displayed (dotted line). The points are colored according to the temperature of acquisition from 21°C to 27°C. The analysis of cilia beating individual parameters was performed on the same records. **C.** Cilia beat frequency (CBF). **D**. Percentage of the video area covered by cilia beating (movement). **E.** Synchronization defined as the surface size below which cilia beat in a homogeneous way ($\lambda^2$). **F** Dispersion of cilia beat orientation. One point per video acquisition. Mean ± SEM. Mann-Whitney tests, * $p<0.05$, **$p<0.01$, ***$p<0.001$, ****$p<0.0001$.

**Figure 3**. **Effects of LXB4 and RvE1 on cilia beat phenotype.** LXB4 (10nM) and RvE1 (10nM) or vehicle (ethanol) was added, 24h after mucus accumulation, into the basolateral compartment of culture inserts from 4 (for LXB4 experiment) and 5 (for RvE1 experiment) CF patients. Video were acquired at different times after LXB4 exposure: at 1h (35 videos with vehicle and 16 with LXB4), at 24h (39 with vehicle and 26 with LXB4) and at 48h (20 with vehicle and 15 with LXB4). Similarly, videos were acquired at different times after RvE1exposure: at 1h (58 with vehicle and 46 with RvE1), at 24h (59 with vehicle and 49 with RvE1) and at 48h (39 with vehicle and 32 with RvE1). Scatter plot on the first two PC is displayed, each point corresponding to one video acquisition. PERMANOVA statistical tests were performed to compare cilia beat phenotype (centroid) recorded on CF hNEC exposed to the vehicle (blue) and CF hNEC exposed to SPMs (green); with p-values >0.5 considered as not significant (NS). PCA analysis from non-CF samples exposed to the vehicle is displayed in grey.

**Figure 4. Distinct effects of RvE1 on hNEC from two F508del homozygous donors**. PCA analysis (PC1 and PC2) corresponding to hNEC from CF1 (**A**) and from CF2 patients (**C**) treated with vehicle and RvE1 (10nM), as well as the non-CF hNEC beating phenotypes. Times correspond to the RvE1 exposure treatment duration. Corresponding values of cilia beating frequency (CBF), synchronization ($\lambda^2$) and orientation for the 2 groups (untreated and RvE1 treated samples) over time for hNEC from CF1 patient (**B**) and from CF2 patient (**D**). Mean ± SEM; * $p<0.05$, **$p<0.01$,***$p<0.001$,****$p<0.0001$. PERMANOVA statistical tests were performed to compare cilia beat phenotype (centroids) recorded on CF hNEC exposed to the vehicle (blue) and CF hNEC exposed to RvE1 (green). Mean ± SEM, * $p<0.05$, ****$p<0.0001$.

**Figure 5. Role of the TMEM16A in the effect of RvE1 on the ASL height and the cilia beating phenotype**. **A**. Representative orthogonal views of confocal microscopy z-stacks showing the ASL for a CF sample in control condition (vehicle) and treated with RvE1 (10nM). ASL is labelled using

dextran Texas red and cells using calcein green. **B**. Quantification of ASL height for CF hNEC from 3 donors (F508del homozygous), in control condition, after RvE1(10nM) treatment and after a combination of RvE1 and Ani 9 (10µm), a specific inhibitor of the TMEM16A channel added 30 minutes before RvE1. The dashed line at 7µm illustrates the ASL height measured in non-CF hNEC. **C**. The role of TMEM16A on cilia beating was investigating by treating the hNEC from 2 CF donors (F508del homozygous), with Ani9 30min before the addition of RvE1. Points and ellipses corresponding to CF vehicle (blue), RvE1 10nM (green) and Ani9 + RvE1 10nM (purple) groups. Times correspond to treatment duration. PERMANOVA tests were performed for Ani9+RvE1 versus control (vehicle). Statistical tests were performed by including all principal components. **D**. Corresponding values of CBF, synchronization and orientation for the 2 groups over time. Mean ± SEM; * $p<0.05$, **$p<0.01$.

**Figure 6. RvE1 modified mucins secretion and structure**. **A.** Representative confocal projection images of MUC5AC (green) and MUC5B (red) secreted apically by non-CF and CF hNECs cultures treated with vehicle solution or RvE1(10nM) ± Ani9 (10µM). Culture inserts were fixed after 48h of treatment. Orthogonal views (Z) of MUC5AC. White scale bar = 10µm. **B.** Average thickness of MUC5AC structures measured on hNEC from 2 CF donors and 4 non-CF donors (2 culture inserts by condition and 3 to 5 acquisitions). **C.** Mean transcription levels of MUC5B and MUC5AC in untreated hNEC from 8 non-CF donors and 6 CF donors and in hNEC derived from 4 CF donors and treated for 24h with RvE1 (10nM). Transcription levels expressed as delta Ct (Ct of the mucin minus Ct of the housekeeping gene). Mean ± SEM. T-test, *$p<0.05$, ****$p<0.0001$.

**Table 1: Samples donor distribution for cilia beating acquisitions.** Median age in years, 1st and 3rd quartiles in brackets.

**Table 2.** Cilia beating parameters changes over time after mucus removal in CF and non-CF hNEC primary cultures. Values are expressed as Mean ± SEM.

Figure 1

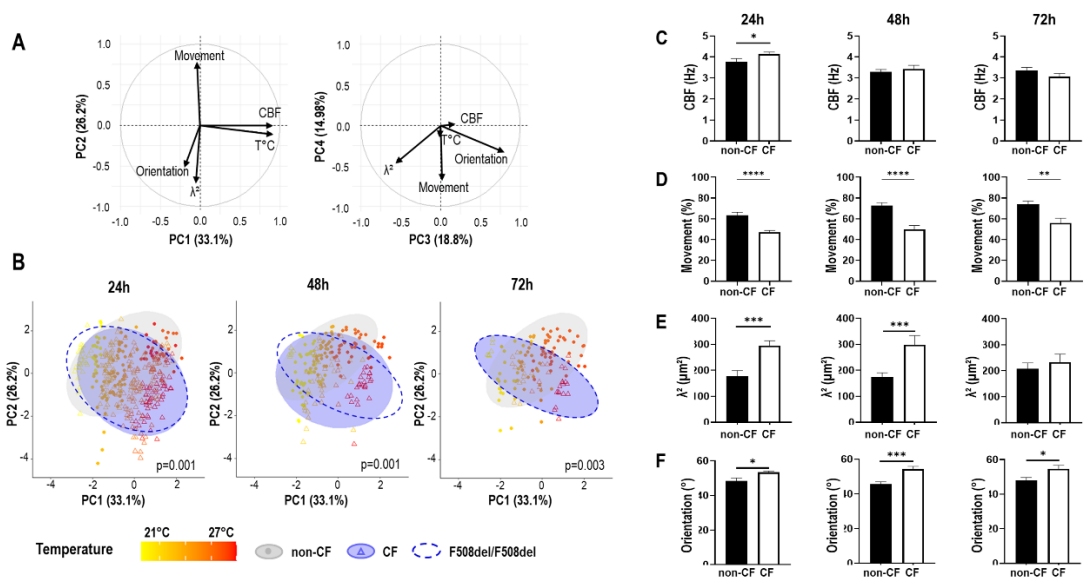

Figure 2

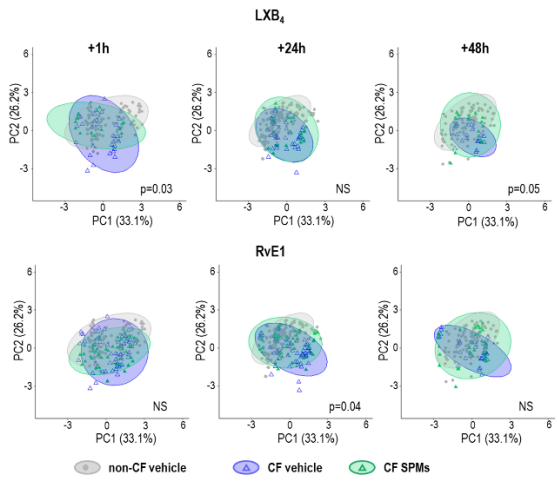

Figure 3

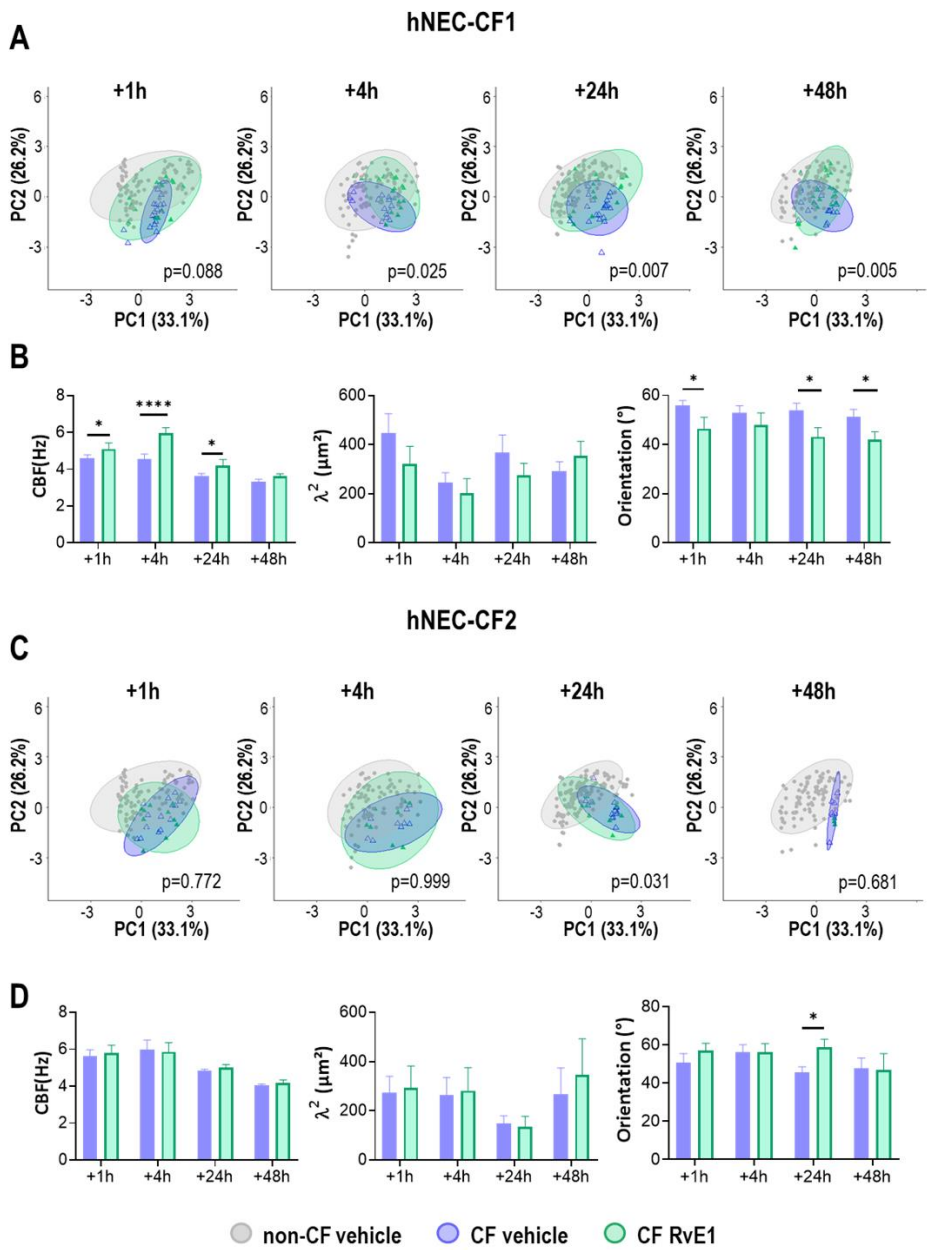

Figure 4

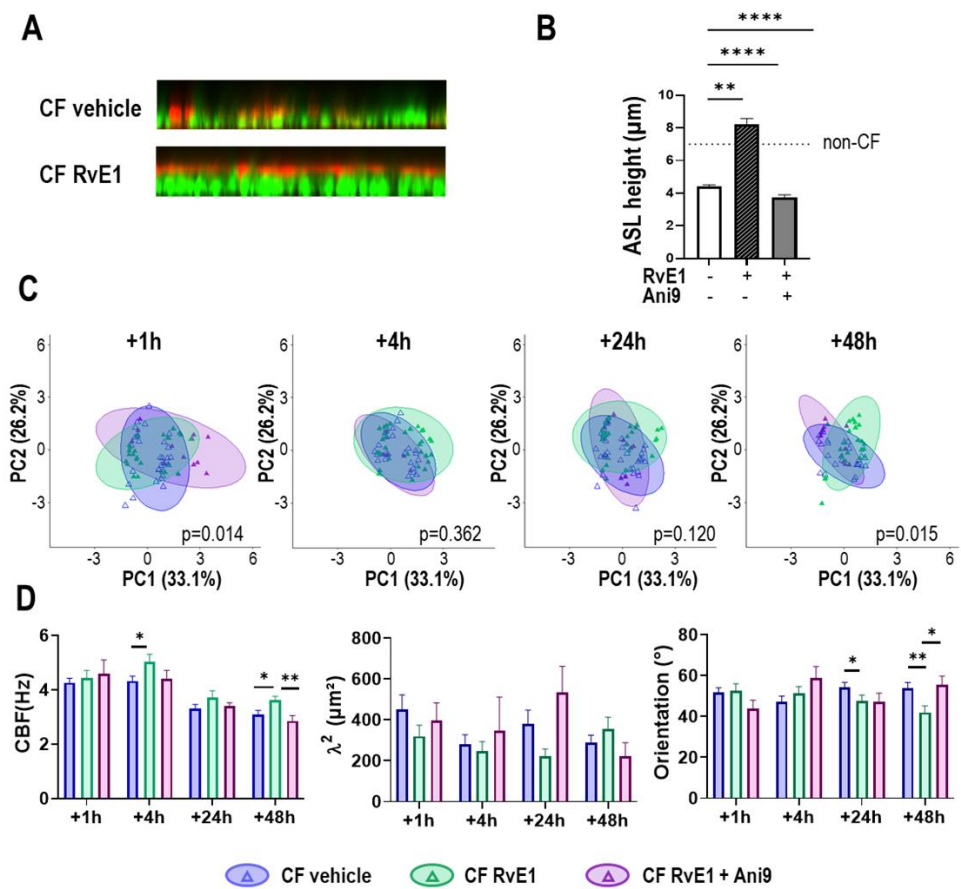

Figure 5

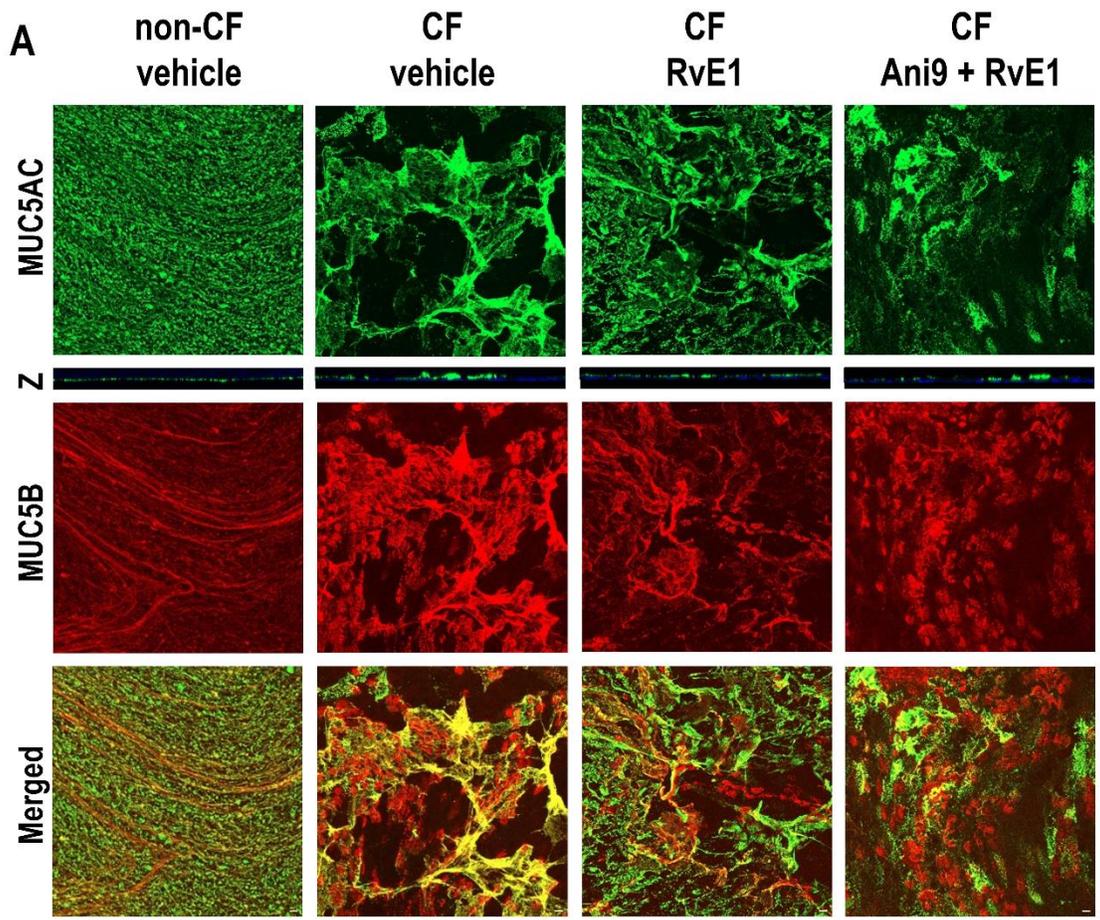
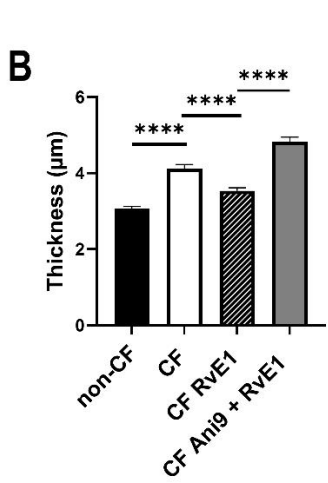
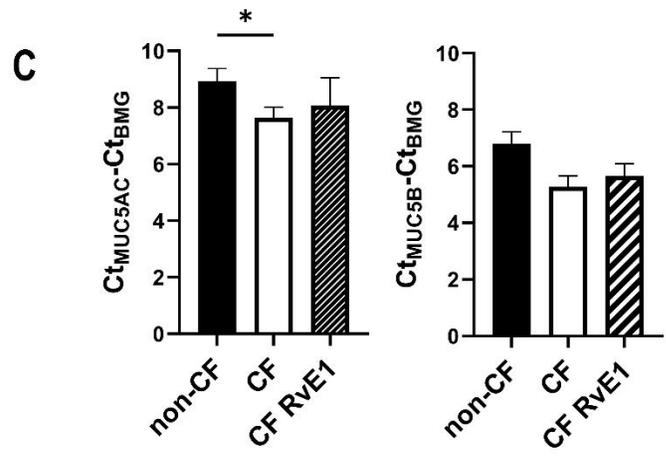

Figure 6

| CFTR genotype | Number | Age | Gender |
|---|---|---|---|
| WT | 9 | 48 (39,53) | 4F/5M |
| F508del/F508del | 4 | 26,25 (17.6, 35.5) | 2F/2M |
| F508del/Y1092X | 1 | 31 | 1M |
| F508del/1717G-A | 1 | 36 | 1F |
| **Total** | 15 | - | 7F/8M |

Table 1

| | | 24h | 48h | p value (48h vs 24h) | 72h | p value (72h vs 24h) |
|---|---|---|---|---|---|---|
| **$\lambda^2$ (µm²)** | non-CF | 178 ± 21 | 175 ± 16 | 0.47 | 208 ± 22 | 0.06 |
| | CF | 295 ± 19 | 298 ± 35 | 0.53 | 232 ± 31 | 0.55 |
| **Orientation (°)** | non-CF | 48.5 ± 1.5 | 45.6 ± 1.4 | 0.12 | 48.1 ± 1.6 | 0.77 |
| | CF | 53.3 ± 0.8 | 54.3 ± 1.6 | 0.46 | 54.7 ± 2 | 0.62 |
| **CBF(Hz)** | non-CF | 3.8 ± 0.1 | 3.2 ± 0.1 | 0.0076 | 3.3 ± 0.1 | 0.0007 |
| | CF | 4.1 ± 0.1 | 3.4 ± 0.1 | 0.04 | 3.1 ± 0.1 | 0.0001 |

Table 2